\documentclass[aps,prl,reprint,groupedaddress,amsmath,amssymb,longbibliography]{revtex4-2}

\usepackage{graphicx}
\usepackage{dcolumn}
\usepackage{bm}
\usepackage{xcolor}
\usepackage{braket}

\begin{document}

\title{Angular dependence of the Wigner time delay upon strong field ionization \\from an aligned p-orbital}

\author{D. Trabert$^1$} 
\email{trabert@atom.uni-frankfurt.de}
\author{N. Anders$^1$}
\author{A. Geyer$^1$}
\author{M. Hofmann$^1$}
\author{M. S. Sch\"offler$^1$}
\author{L. Ph. H. Schmidt$^1$}
\author{T. Jahnke$^2$}
\author{M. Kunitski$^1$}
\author{R. D\"orner$^1$}
\email{doerner@atom.uni-frankfurt.de}
\author{S. Eckart$^1$}
\email{eckart@atom.uni-frankfurt.de}

\affiliation{$^1$Institut f\"ur Kernphysik, Goethe-Universit\"at, Max-von-Laue-Str. 1, 60438 Frankfurt am Main, Germany \\
$^2$ European XFEL, Holzkoppel 4, 22869 Schenefeld, Germany
}

\date{\today}
\begin{abstract}
We present experimental data on the strong-field ionization of the argon dimer in a co-rotating two-color (CoRTC) laser field. We observe a sub-cycle interference pattern in the photoelectron momentum distribution and infer the Wigner time delay using holographic angular streaking of electrons (HASE). We find that the Wigner time delay varies by more than 400 attoseconds as a function of the electron emission direction with respect to the molecular axis. The measured time delay is found to be independent of the parity of the dimer-cation and is in good agreement with our theoretical model based on the ionization of an aligned atomic p-orbital.
\end{abstract}

\maketitle
 
The Wigner time delay \cite{Wigner1954} is a well-established concept in quantum mechanics and is frequently employed to quantify time scales in photoionization processes. Photoionization time delays for single-photon and strong-field ionization of non-isotropic systems strongly depend on the emission direction of the electron \cite{Hockett2016,vos2018orientation,Rist2021,Holzmeier2021,Trabert2021,Guo2022}. However, due to experimental challenges, measurements which do not average over the emission angle are scarce and were only done with molecules. Here, we report on a joint experimental and theoretical study which quantifies the Wigner time delay for strong-field ionization of an aligned atomic p-orbital and find variations of more than 400 attoseconds as a function of the electron emission angle.

For single-photon ionization there are three classes of experimental techniques that are frequently used to access time delays: attosecond streaking \cite{Nagele2011}, RABBITT (Reconstruction of Attosecond Harmonic Beating By Interference of Two-photon Transitions) \cite{Paul2001,Muller2002,Klunder2011,Bharti2021,Gong2022a}, and most recently an approach employing electron angular emission distributions in the molecular frame \cite{Rist2021,Holzmeier2021}. In the strong-field regime, holographic angular streaking of electrons (HASE) has been used in order to measure Wigner time delays \cite{EckartHASE2020,Eckart2020,Trabert2021} and is the strong-field analogon of the RABBITT technique. HASE employs a $\omega$-2$\omega$ bicircular co-rotating two-color (CoRTC) laser field comprised of a strong $2\omega$-component with a central wavelength of 395\,nm  and a weak $\omega$-component with a central wavelength of 790\,nm. This tailored two-color laser field gives rise to a two-path interference that leads to the formation of sidebands (SB) and a strong modulation of the above-threshold ionization (ATI) peaks which has been demonstrated theoretically and experimentally \cite{Feng2019,EckartHASE2020,Eckart2020,Han2018,Ge2019,Trabert2021}. 
The electron momentum distribution shows a characteristic alternating half-ring pattern. The central observable in HASE experiments is the rotation angle of the alternating half-ring pattern. This rotation angle allows one to infer the phase gradient on the electronic wave packet in initial momentum space perpendicular to the tunnel exit $\Phi_{\mathrm{init}}'$ \cite{Liu2016}. Studying the ionization of H$_{2}$ it was shown that the alternating half-ring structure carries information about the shape of the molecular orbital and allows for the measurement of the corresponding contribution to the Wigner time delay \cite{Trabert2021,EckartHASE2020}. For the strong-field ionization of molecular hydrogen, the Wigner time delays varied on the order of 40 attoseconds as a function of the emission angle relative to the molecular axis \cite{Trabert2021}. 

\begin{figure}[b]
\includegraphics[width=\columnwidth]{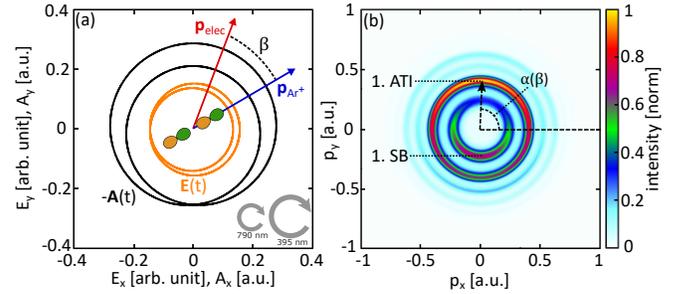}
\caption{\label{fig1} Illustration of two aligned p-orbitals of an argon dimer in a laser electric field and the measured electron momentum distribution. (a) depicts the co-rotating two-color (CoRTC) field $\mathbf{E}(t)$ and the corresponding negative vector potential $-\mathbf{A}(t)$ that are used in the experiment. The helicity of the two circularly polarized pulses is indicated by the grey arrows. The angle $\beta$ is defined as the angle between the electron momentum vector and the momentum vector of the Ar$^{+}$ ion in the plane of polarization. (b) shows the measured electron momentum distribution in the plane of polarization, which shows the characteristic alternating half-ring structure. The first above-threshold ionization (ATI) and sideband (SB) peaks are indicated. For each peak, an angle $\alpha$ can be determined that indicates the most probable electron emission angle in the laser field frame of reference (here $\alpha$ is indicated for the first ATI peak).}
\end{figure}

In this Letter, we will investigate the van-der-Waals-bound argon dimer using the same approach as in Ref. \cite{Trabert2021}. The argon dimer has a much larger internuclear distance of 3.76\,\AA\ in the ground state \cite{Herman1988} as compared to the two protons in the H$_{2}$ molecule, which are separated by only 0.74\,\AA . Interestingly, the dimer only dissociates with a significant kinetic energy release (KER) of the dissociation products if the created electron hole in both of the atoms has the shape of a p-orbital that is aligned along the molecular axis. Previous works have demonstrated the experimental discrimination of dimer cations with gerade and ungerade parity by means of the KER \cite{Sann2016,Kunitski2019}. It was found, for example, that the two-center interference fringes in the molecular-frame photoelectron momentum distribution \cite{Cohen1966} are inverted upon selecting the gerade parity hole state. As we show below, the Wigner time delay is determined solely by the aligned atomic p-orbital and is insensitive to dimer properties such as the parity of the dimer orbital. Thus, the argon dimer Ar$_{2}$ allows us to experimentally access an aligned atomic orbital, which typically poses serious experimental challenges for isolated atoms \cite{Fleischer2011,Fechner2014}.  

Fig. \ref{fig1}\,(a) shows the time-dependent electric field $\mathbf{E}(t)$ and the corresponding negative vector potential $-\mathbf{A}(t)$ of the CoRTC field which was used throughout this Letter. To generate the CoRTC field, laser pulses at a wavelength of 790\,nm pass a beam splitter. Here light at 790\,nm is referred to light at the fundamental frequency $\omega$. A 200\,\textmu m $\beta$-barium borate crystal is used to frequency-double one of the pulses. The intensity and polarization of the $\omega$ and the $2\omega$ pulses can be set independently. A piezo delay stage allows us to ensure temporal overlap and to control the relative phase of the two laser pulses. The two laser pulses are set to have circular polarization with identical helicity. The combined laser electric field thus forms a CoRTC field and is focused by a spherical mirror (focal length $f=60$\,mm) onto a cold supersonic jet of argon resulting in an intensity [peak electric field] of $5.9\times 10^{13}$\,W/cm$^{2}$ [$E_{2\omega}=$ 0.029\,a.u.] for the $2\omega$, and $3.4\times 10^{11}$\,W/cm$^{2}$ [$E_{\omega}=$ 0.0022\,a.u.] for the $\omega$ component of the CoRTC field. The cold gas jet is generated by expanding argon gas through a 30\,\textmu m nozzle into vacuum. Using a backing-pressure of 2.1\,bar, a small fraction of the target gas condensates during the expansion, yielding a small fraction of dimers in the jet, as well. After the ionization of a dimer, electrons and ionic fragments are accelerated in a spectrometer by static electric (26.76\,V/cm) and magnetic (14.2 Gauss) fields towards position- and time-sensitive detectors of a cold target recoil ion momentum spectroscopy (COLTRIMS) reaction microscope \cite{Dorner2000}. Each detector is comprised of a double-stack of micro channel plates and a hexagonal delay-line anode \cite{Jagutzki_2002}. The ion [electron] arm of the spectrometer has a length of 17\,cm [30.6\,cm]. The three-dimensional momentum vectors of all charged fragments that emerge from an ionization event are measured in coincidence. In our current work, we study single ionization and subsequent dissociation of the argon dimer. To this end, $\beta$ denotes the angle between the final electron momentum and the momentum vector of the Ar$^{+}$ ion in the plane of polarization, which is the xy-plane (see Fig. \ref{fig1}\,(a)). Thus, our experimental setup allows for the measurement of the electron momentum, the kinetic energy release and $\beta$ in coincidence for each dissociation event.

Fig. \ref{fig1}\,(b) shows the momentum distribution in the plane of polarization of all electrons measured in coincidence with Ar$^{+}$ ions. The momentum distribution exhibits the well-known alternating half-ring structure comprised of above-threshold ionization (ATI) and sideband (SB) peaks. For each peak, an angle $\alpha$ can be determined, which indicates the maximum of the corresponding angular distribution in the laser field frame of reference. In the following, the rotation angle of the alternating half-ring structure $\alpha$ will be analyzed as a function of $\beta$ for three different dissociation channels of the argon dimer. 

\begin{figure}[b]
\includegraphics[width=1 \columnwidth]{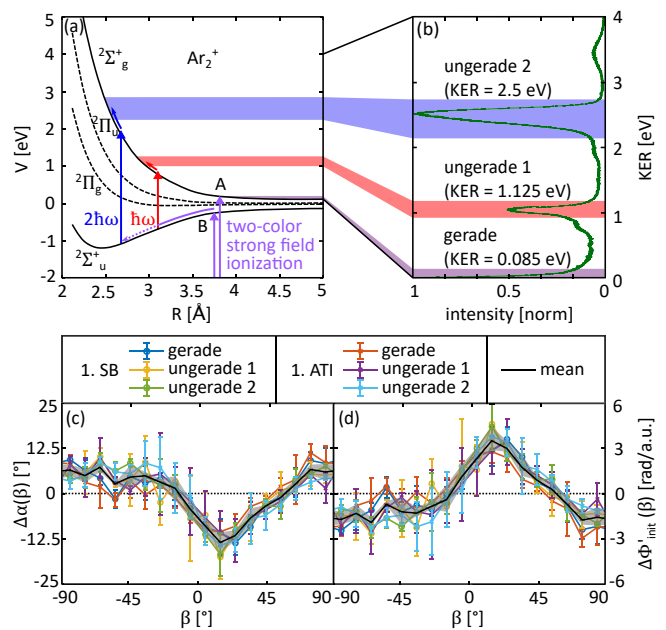}
\caption{\label{fig2} Parity-resolved ionization pathways of the argon dimer and HASE-observables. (a) Potential energy curves of the $^{2}\Sigma_{g}^{+}$ and the $^{2}\Sigma_{u}^{+}$ state of the Ar$_{2}^{+}$ ion are populated by two-color strong field ionization (purple arrow). When reaching the $^{2}\Sigma_{g}^{+}$ state directly (pathway A), immediate dissociation occurs. The other purple arrow (pathway B) indicates the starting point of wave packets on the $^{2}\Sigma_{u}^{+}$ potential energy curve that propagate to smaller internuclear distances until one-photon transitions to the dissociative $^{2}\Sigma_{g}^{+}$ state become resonant. (b) Measured kinetic energy release (KER) for the dissociation of the argon dimer via the pathways depicted in (a). The three KER-regions of interest are highlighted and labeled by 'gerade', 'ungerade 1' and 'ungerade 2'. (c) shows the measured result for $\Delta\alpha\left(\beta\right)$ for the three breakup channels and the two energy peaks (1. SB and 1. ATI) highlighted in Fig. \ref{fig1}\,(b). The black line shows the average of the data points that are shown in color. (d) depicts the experimentally obtained phase gradient $\Delta\Phi_{\mathrm{init}}'$ as a function of $\beta$. Error bars, indicated by the grey area, in (c) and (d) show the statistical error only.}
\end{figure}

Fig. \ref{fig2}\,(a) shows the potential energy curves of the involved ionic states and the possible pathways for the reaction $\mathrm{Ar_{2}}\rightarrow\mathrm{Ar^{+}}+\mathrm{Ar}+\mathrm{e^{-}}$. The curves are taken from Ref. \cite{Stevens1977}, spin-orbit interaction is neglected. Pathway A directly populates the gerade $^{2}\Sigma_{g}^{+}$ state of the Ar$_{2}^{+}$ ion, leading to a small kinetic energy release (KER) of the direct dissociation as can be seen in Fig. \ref{fig2}(b). Pathway B creates a wave packet on the ungerade $^{2}\Sigma_{u}^{+}$ energy curve that subsequently propagates to smaller internuclear distances. In a second step, resonant transitions mediated by one photon of either the $\omega$ or the $2\omega$ component of the CoRTC field populate the dissociative $^{2}\Sigma_{g}^{+}$ state. This leads to characteristic peaks in the KER distribution which are marked red and blue in Fig. \ref{fig2}(b). 

For each of the highlighted KER-regions, two-dimensional electron momentum distributions similar to Fig. \ref{fig1}(b) can be extracted from our total data set for different values of $\beta$ to obtain $\alpha\left(\beta\right)$. An interval of 10 degrees has been chosen for each value of $\beta$. The changes of $\alpha\left(\beta\right)$ are defined as $\Delta\alpha\left(\beta\right)=\alpha\left(\beta\right)-\alpha_{\mathrm{mean}}$, where $\alpha_{\mathrm{mean}}$ denotes the mean angle $\alpha$ for each energy peak and breakup channel integrated over all values of $\beta$. The analysis and symmetrization procedure to obtain $\Delta\alpha\left(\beta\right)$ is in full analogy to that described in Ref. \cite{Trabert2021}. $\Delta\alpha\left(\beta\right)$ is shown in Fig. \ref{fig2}(c) for the first ATI, the first SB peak and for electrons released from the gerade and from the ungerade orbital separately. Surprisingly, within error bars, $\Delta\alpha\left(\beta\right)$ is identical for all dissociation channels, implying that it is independent of the parity of the ionized molecular orbital. This is a strong indication, that for rare gas dimers, the observable $\Delta\alpha\left(\beta\right)$ depends on atomic rather than on dimer properties. We note that this is a characteristic of this observable and that other observables such as the molecular-frame photoelectron momentum distribution strongly depend on the parity and show a pronounced two-center interference \cite{Kunitski2019}.

Since we find no significant dependence of $\Delta\alpha\left(\beta\right)$ on the dimer orbital parity, for the remainder of this paper we use the mean value of $\Delta\alpha\left(\beta\right)$ averaged over all three breakup channels. This reduces the statistical error for a given value of $\beta$. The mean value of $\Delta\alpha\left(\beta\right)$ averaged over all breakup channels and both electron energy peaks (1. SB and 1. ATI) is shown as black line in Fig. \ref{fig2}(c). The grey area indicates the corresponding statistical error. 

To further analyze the data we use a semi-classical trajectory-based simulation as described in Ref. \cite{Trabert2021} to deduce the phase gradient on the initial momentum perpendicular to the direction of the laser's electric field at the instant of tunneling $\Delta\Phi_{\mathrm{init}}'\left(\beta\right)$ from the measured values of $\Delta\alpha\left(\beta\right)$. Fig. \ref{fig2}\,(d) shows the experimental result for $\Delta\Phi_{\mathrm{init}}'$ as a function of $\beta$. We find a sinusoidal dependence on $\beta$ and values of up to 3\,rad/a.u. The magnitude of the measured phase gradients is significantly larger than in previous works on molecular hydrogen \cite{Trabert2021} and carbon monoxide \cite{Guo2022}.  

\begin{figure}[t]
\includegraphics[width=1\columnwidth]{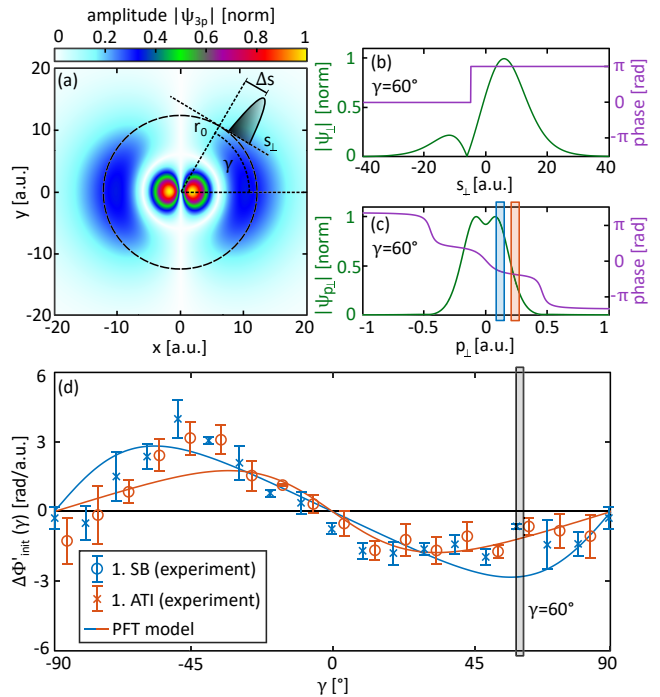}
\caption{\label{fig3}
Microscopic origin of Wigner time delays for the strong field ionization from an atomic p-orbital aligned along the internuclear axis. (a) shows the amplitude $|\Psi_{\mathrm{3p}}|$ of an atomic $3p_{0}$ orbital in position space. According to the partial-Fourier transform (PFT) model, tunneling probes the bound wave function along the spatial component $s_{\perp}$. The angle $\gamma=60^{\circ}$ between $\mathbf{r}_{0}$ and the internuclear axis as well as the position offset $\Delta s$ are illustrated. (b) One-dimensional wave function along $s_{\perp}$ for $\gamma=60^{\circ}$ in position space representation. (c) Momentum space representation of the wave function shown in (b). The most probable initial momentum for the 1. ATI [1. SB] peak is highlighted by the red [blue] box. (d) The measured result for $\Delta\Phi'_{\mathrm{init}}$ is plotted as a function of $\gamma$ for the 1. ATI [1. SB] peak. The data for the two electron energy peaks are shown as red circles [blue marks] after integration over all three breakup channels. The corresponding predictions from the PFT model are illustrated as solid lines. The exemplary case of $\gamma=60^{\circ}$ is highlighted by the grey box. Error bars in (d) show the statistical error only.}
\end{figure}

\begin{figure}[t]
\includegraphics[width=1\columnwidth]{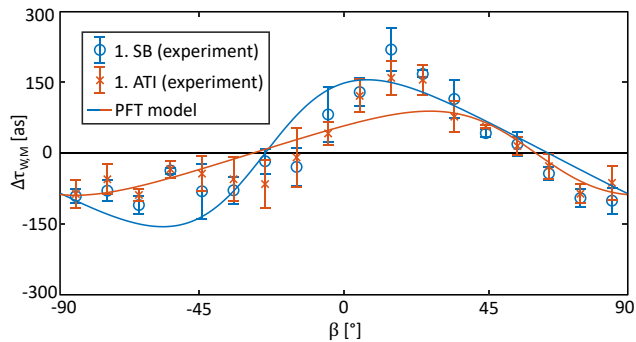}
\caption{\label{fig4} Measured and calculated dependence of the Wigner time delay $\Delta\tau_{\mathrm{W,M}}$ on the electron's emission direction in the molecular frame $\beta$. Values of more than 200 attoseconds are observed. The data points show a sinusoidal dependence on the angle $\beta$ and are in good agreement with the predictions from our PFT model. Error bars show the statistical error only.}
\end{figure}

In a next step, we compare our measured phase gradients $\Delta\Phi_{\mathrm{init}}'\left(\beta\right)$ to the predictions of a simple and instructive approach \cite{Trabert2021} which models the strong-field ionization from a single 3$p_{0}$ orbital aligned along the internuclear axis of the dimer. This procedure is motivated by the independence of $\Delta\Phi_{\mathrm{init}}'\left(\beta\right)$ on the parity of the ionic state as well as by previous theoretic treatments \cite{Kunitski2019} that model the strong-field ionization process of noble gas dimers as the coherent sum of two atomic ionization processes. Fig. \ref{fig3}\,(a) shows the absolute value $|\Psi_{\mathrm{3p}}|$ of the wave function representing a 3$p_{0}$ orbital. We define the angle of the tunnel exit position vector $\mathbf{r}_{0}$ with respect to the dimer axis as $\gamma$. Next, we use that tunneling acts as a filter on the bound wave function in position space in a good approximation \cite{Meckel2008,Arissian2010,Kang2020}. Hence, for most values of $\gamma$, the spatial structure of a p-orbital at the distance of the tunnel exit $r_{0}$ results in an effective shift $\Delta s$ of the tunneling wave function perpendicular to the tunnel direction. In particular, we apply the partial-Fourier transform (PFT) model \cite{Murray2010,Liu_2016} by evaluating the one-dimensional position space wave function $\Psi_{\perp}$ of the single 3$p_{0}$ orbital along the spatial component $s_{\perp}$ at the distance of the tunnel exit $r_{0}$ for different values of $\gamma$. Fig. \ref{fig3}(b) shows the resulting slice of the position space wave function for the exemplary case of $\gamma=60^{\circ}$. Here, the amplitude is clearly shifted to larger values of $s_{\perp}$ and the phase exhibits a phase jump of $\pi$, as expected for a p-orbital. The Fourier transform of each one-dimensional slice yields the corresponding initial momentum space representation $\Psi_{p_{\perp}}$ of the electronic wave packet at the tunnel exit as a function of $p_\perp$. Here, $p_\perp$ is the momentum perpendicular to the light's electric field at the instant of tunneling. The released wave packet in momentum space is schematically shown in Fig. \ref{fig3}\,(c) for the case of $\gamma=60^{\circ}$. The phase of $\Psi_{p_{\perp}}$ has a varying negative slope which corresponds to the expected value of $\Delta\Phi_{\mathrm{init}}'\left(\gamma\right)$ for a specific value of initial momentum $p_{\mathrm{initial}}$ perpendicular to the tunnel exit (see Eq. (5) in Ref. \cite{Trabert2021}). The initial momentum $p_{\mathrm{initial}}$ of 0.12 [0.24]\,a.u. corresponds to the 1. SB [1. ATI] peak and is highlighted by a blue [red] box in Fig. \ref{fig3}\,(c). The values of $\Delta\Phi_{\mathrm{init}}'\left(\gamma\right)$ expected from our model for the 1. ATI as well as the 1. SB peak are shown in Fig. \ref{fig3}\,(d) as a function of $\gamma$.  The curves have a sinusoidal shape and are zero for $\gamma=0^{\circ},\pm 90^{\circ}$ due to the symmetry of $\Psi_{\perp}$ for these cases. The theoretical model is in very good agreement with the corresponding experimental values for $\Delta\Phi_{\mathrm{init}}'\left(\gamma\right)$ (see Fig. \ref{fig3}\,(d)). For this comparison, the angle $\gamma$ in position space was inferred from the measured angle $\beta$ in momentum space using the relation $\gamma=\beta+90^{\circ}+\kappa$. Coulomb interaction between the electron and the parent ion after tunneling results in a non-zero rotation of the electron momentum distribution by the angle $\kappa$ that is on the order of 20$^{\circ}$ to 30$^{\circ}$. The experimental data are averaged over all three breakup channels. The value of $p_{\mathrm{initial}}$ and the angle $\kappa$ for each measured final electron momentum have been determined by a semi-classical trajectory-based model in full analogy to Ref. \cite{Trabert2021}, assuming a circularly polarized $2\omega$ light field with a peak electric field of 0.029\,a.u. The exemplary case of $\gamma=60^{\circ}$ is highlighted by the grey box in Fig. \ref{fig3}\,(d) and shows the two negative phase gradients of varying slope. $\Delta\Phi_{\mathrm{init}}'$ can be approximately interpreted as the position offset $-\Delta s$. Hence, our measured data provides direct information about the geometry of the bound electronic orbital in position space representation. This correspondence would be exact if $\Delta\Phi_{\mathrm{init}}'$ was independent of $p_\perp$.

In the last section of this Letter, we present the angular dependence of the Wigner time delay $\Delta\tau_{\mathrm{W,M}}$ extracted from the previously discussed experimental and theoretical results using Eq. (6) from Ref. \cite{Trabert2021}. In Fig. \ref{fig4} the measured and the calculated values of $\Delta\tau_{\mathrm{W,M}}$ are shown as a function of $\beta$. The sinusoidal shape and the good agreement between experiment and PFT model are preserved compared to Fig. 3. Overall, there is a variation of more than 400\,as, which is one order of magnitude larger than what has been observed for H$_{2}$ \cite{Trabert2021}. 

In conclusion, we have analyzed a sub-cycle photoelectron interference upon the strong-field ionization of the argon dimer using a co-rotating two-color (CoRTC) laser field. The interference is analyzed using HASE to infer the phase gradients of the electronic wave function at the tunnel exit in momentum space as well as the corresponding changes of the Wigner time delay for the electronic wave function in the continuum. The measured phase gradients and Wigner time delays are in good agreement with the expectation for the strong-field ionization of an aligned atomic 3$p_{0}$ orbital. This leads to a microscopic understanding of the ionization process and reveals how the complex valued continuum wave function's properties depend on the electron's bound state for a p-orbital, which is a general finding in the strong-field regime. We report that the Wigner time delay varies by more than 400 attoseconds as a function of the electron emission direction with respect to the molecular axis. Our work paves the way for future approaches to characterize the phase of the continuum wave function of liberated electrons from more complex atomic and molecular orbitals.

\section{Acknowledgments}
\normalsize
The experimental work was supported by the DFG (German Research Foundation). S.E. acknowledges funding of the DFG through Priority Programme SPP 1840 QUTIF. We acknowledge fruitful discussions with Simon Brennecke and Manfred Lein.

\end{document}